\documentclass[%
 reprint,
 amsmath,amssymb,
 aps,
]{revtex4-2}
\usepackage[nointlimits]{esint}
\usepackage[utf8]{inputenc}
\usepackage[T1]{fontenc}
\usepackage{mathptmx}
\usepackage{etoolbox}
\usepackage{xcolor}
\usepackage{xpatch}
\usepackage{ulem}
\usepackage{graphicx}
\usepackage{dcolumn}
\usepackage{bm}
\usepackage{algorithm}
\usepackage[mathlines]{lineno}

\makeatletter

\begin{document}

\preprint{APS/123-QED}

\title{Quadrupole topological behavior of elastic waves in two-dimensional square lattices with nonsymmorphic symmetries}

\author{Yijie Liu}
\affiliation{School of Civil and Transportation Engineering, Guangzhou University, 230 Wai Huan Xi Road, Guangzhou, 510006, China }

\author{Yuyang Chen}%
\affiliation{School of Civil and Transportation Engineering, Guangzhou University, 230 Wai Huan Xi Road, Guangzhou, 510006, China }

\author{Zhaoyang Guo}%
\affiliation{School of Civil and Transportation Engineering, Guangzhou University, 230 Wai Huan Xi Road, Guangzhou, 510006, China }

\author{Zhi-Kang Lin}
\address{School of Physical Science and Technology, and Collaborative Innovation Center of Suzhou Nano Science and Technology, Soochow University, Suzhou 215006, China}

\author{Di Zhou}
\email{Corresponding author: dizhou@bit.edu.cn}
\affiliation{Key Lab of Advanced Optoelectronic Quantum Architecture and Measurement (MOE), School of Physics, Beijing Institute of Technology, Beijing 100081, China}%

\author{Feng Li}
\email{Corresponding author: phlifeng@bit.edu.cn}
\affiliation{Key Lab of Advanced Optoelectronic Quantum Architecture and Measurement (MOE), School of Physics, Beijing Institute of Technology, Beijing 100081, China}%
 
\author{Ying Wu}
\email{Corresponding author: yingwu@njust.edu.cn}
\affiliation{School of Physics, Nanjing University of Science and Technology, Nanjing 210094, China}

\date{\today}

\begin{abstract}
We investigate a novel higher-order topological behavior in elastic lattices characterized by nonsymmorphic symmetries. In the theoretical spring-mass lattice, altering the vertex mass allows for fine-tuning of the topological features within the bandgap. We analyze the quadrupole topological behavior in square lattices with nonsymmorphic symmetries using nested Wannier bands. Beyond second-order topological metamaterials, a single-phase topological configuration promotes energy localization at the corners due to a non-zero relative quadrupole moment. Our findings are validated through experimental observations of higher-order topological corner states, which show excellent agreement with simulated results and theoretical predictions. Additionally, the elastic lattices in the self-similar system exhibit fractal higher-order topological behaviors, revealing numerous topological edge and corner states. The self-similar lattice also demonstrates enhanced energy localization, with the number of topological states showing a linear correlation to the corner dimension. This study provides new insights into elastic higher-order topological insulators and inspires innovative strategies for simulating topological elastic materials.
\end{abstract}

\maketitle


\section{Introduction}

The topological phases of matter, originating in condensed matter physics, has significantly enhanced research in photonics, acoustics, and mechanics. Following the development of topological interface modes in different dimensions, a series of application strategies are presented, such as multiband transmission \citep{chen2019valley,huang2021resonant,yuan2021multi}, active control \citep{you2021reprogrammable,zhang2018topological,wang2020tunable,chen2023valley} and energy harvesting \citep{wu2023enhanced,li2023acoustic,chen2019topological,liu2021tunable}. Recently, the concept of high-order topological insulators (HOTI)  that go beyond the principle of bulk-edge correspondence has been proposed. Compared to first-order topological insulators, HOTIs can lead to corner \citep{ezawa2018higher,luo2023efficient,wang2023topological} or hinge\citep{wei20213d,schindler2018higher,zhang2019dimensional} states lower than the dimension of the system boundary. For instance, second-order topological insulators in two-dimensional (2D) systems exhibit 1D gapped edge states and 0D corner states. Given the capacity to demonstrate various topological phenomena across several dimensions, HOTIs provide distinct advantages and potential for multifunctional sensing devices.

Currently, HOTIs are classified into two main categories in mechanics. One type is the Wannier-type HOTIs, characterized by the position of the Wannier center and topological invariants \citep{fan2019elastic,wang2020higher,wang2021elastic,wu2021chip,zheng2022higher,liu2023second}. Inspired by the construction framework of the first-order topological insulators, topological corner states have been demonstrated to exist in the Quantum Spin Hall\citep{huo2019edge,fan2019elastic} and Quantum Valley Hall \citep{an2022second} systems. However, in addition to requiring superior waveguide efficiency, modern communication devices often demand multiband operation or switchable functionality. To address the issue of flexible access to the corner state, a series of strategies are being developed, such as topological optimization \citep{zheng2023switchable,chen2023topology}, piezoelectricity \citep{ma2023tuning,yiReconfigurable} and multi-mode \citep{zhang2022helical}. Moreover, the mechanical lattice composed of connecting stiffness and vertex mass is also considered a paradigm for constructing zero-dimensional corner states. Utilizing the theoretical framework evolved from the Su-Schrieffer-Heeger (SSH) model \citep{wu2020plane,zhang2023higher,zhang2023active}, the relationship between topological phase and topological indices in rectangular, kagome, square, and hexagonal lattices is systematically supplemented \citep{duan2023numerical}. A series of interesting phenomena extend the design scheme of the higher-order topological corner state, including rainbow trapping effect \citep{zhou2023visualization}, twisted lattices\citep{Zhang2023Twist}, and high frequency\citep{hong2023high}.

The other type of HOTI is the multipole insulator with quantized multipole polarization \citep{ni2020demonstration,he2020quadrupole,bao2019topoelectrical,dutt2020higher}. Benalcazar initially extended the concept of electric polarization to higher electric multipole moments in the tight-binding model \citep{benalcazar2017quantized,benalcazar2017electric}. The nested Wannier band method can be used to distinguish the high-order insulating phases. Soon, the realization method of four-level subtopological insulators was found in various classical wave fields, with the critical step being the simultaneous action of positive and negative jump coupling. Both optical and acoustic platforms have successfully simulated positive and negative jump coupling, realized by introducing additional path lengths \citep{mittal2019photonic} in nanophotonics silicon ring resonator and coupling direct and cross-connected air cavities \citep{qi2020acoustic}, respectively.

Similarly, mechanical quadrupole HOTIs are also constructed using different curved beam coupled panels \citep{serra2018observation} in a tetragonal lattice. However, only a few positive and negative jump schemes have been proposed in the mechanical fields, resulting in the high working frequency of elastic quadrupole HOTIs. The photonic platform has recently proposed novel design schemes for the quadrupole topological phase \citep{zhou2020twisted,kang2024multiband}, but these are limited to integer-dimension systems as opposed to fractional dimension in fractal systems. As a classical configuration of a non-integer dimension system, the fractal topological behaviors are predominantly conducted within the domains of photonic \citep{ren2023theory,xie2021fractal} and acoustic waves \citep{zheng2022observation, LI20222040}. Given the significance of having rich topological corner and edge states for obtaining precise and reliable elastic wave sensing and transmission, mechanical quadrupole HOTIs provide significant potential for signal manipulation and require immediate innovation.

In this work, we propose 2D square mechanical lattices with nonsymmorphic symmetries to demonstrate the high-order topological behavior of elastic waves. Non-symmorphic symmetries combine a fractional lattice translation with either a mirror reflection (glide plane) or a rotation (screw axis), leading to band folding with crossings at the Brillouin zone boundaries. These crossings are protected against hybridization. The efficacy of characterizing the topological phases of matter with discrete systems has been unequivocally validated \citep{liu2024observation,gao2024acoustic}. Using a spinless SSH model, the nontrivial orbital-induced topological insulators and wave interactions between multiple orbitals are experimentally demonstrated \citep{gao2023orbital}. The Stiefel-Whitney classes in acoustic wave systems are also explored in discrete systems \citep{xiang2024demonstration}. Furthermore, we analyze the wave propagation in fractal mechanical lattice with quadrupole topology. The variations in vertex mass within the spring-mass model are employed to identify the high-order topological band gap. The ''anomalous quadrupole topology" is confirmed for the system by applying the derived effective Hamiltonian into the approach of the nested Wannier band, in which the non-trivial relative quadrupole moment is quantized to $\frac{1}{4}-(-\frac{1}{4})=\frac{1}{2}$. We construct the higher-order self-similar structure, in which we numerically and experimentally demonstrate the edge states and corner states. Remarkably, the finite element simulation and experiment exhibit excellent concordance with the theoretical analysis. Compared with the behavior of acoustic plates in classical Ref. \citep{LI20222040}, the similarities and differences between fractal and self-similarity structures are expounded. In the Sierpi$\rm \acute{n}$ski carpet, the fractal dimension in physics can be intuitively reflected by the limited number of states calculated through the box-counting method. However, self-similar structures are more advantageous in terms of energy local performance, which yields valuable insights for sensing and energy harvesting devices.

\section{Topological properties in 2D square spring-mass lattices}

\begin{figure*}
\includegraphics[scale=0.55]{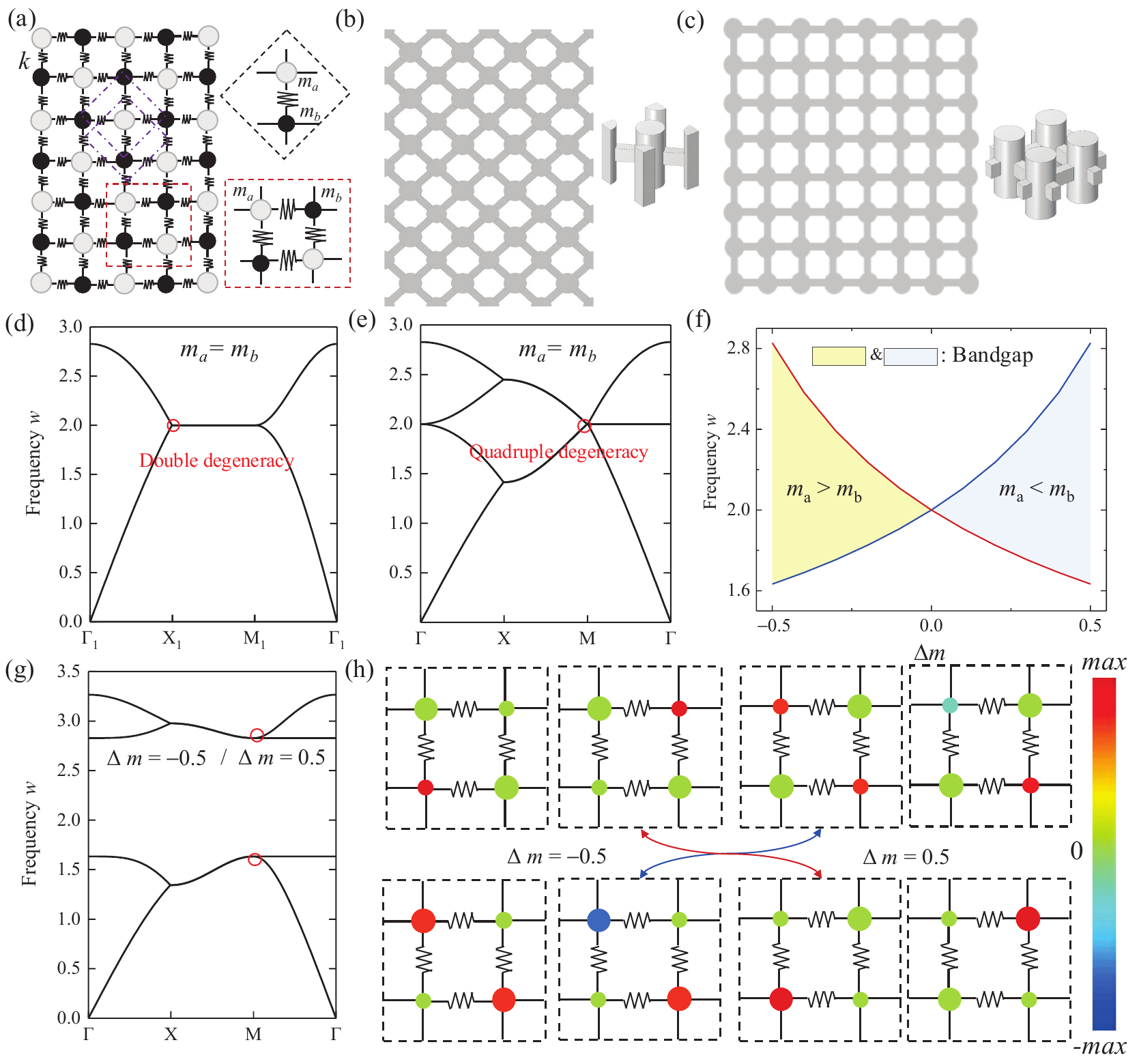}
\caption{\label{fig1a} (a) The finite array form of a 2D spring-mass square lattice, where the area framed by rhombus and rectangle box represents the primitive unit cell with minimum volume and the extended cell, respectively; (b) Top view of the lattice structure consisting of the continuum primitive cell, which corresponds to the rhombus dash box; (c) Top view of the lattice structure consisting of the continuum extended cell, which corresponds to the rectangle dash box; (d)-(e) The band structure of the primitive and expanded unit with $m_{a}$ = $m_{b}$ = 1; (f) Variation curve as a function of $\Delta m$, where the red and blue solid lines show the variation of frequency at M point in Brillouin zone associated with bandgap; (g) The band structure of unit cell with $\Delta m =-0.5$ and $0.5$, where the frequency of bandgap ranges from 1.63 to 2.83; (h) Normalized z displacement field at M point associated with bandgap when $\Delta m =-0.5$ and $0.5$. }
\end{figure*}

\begin{figure}
\includegraphics[scale=0.42]{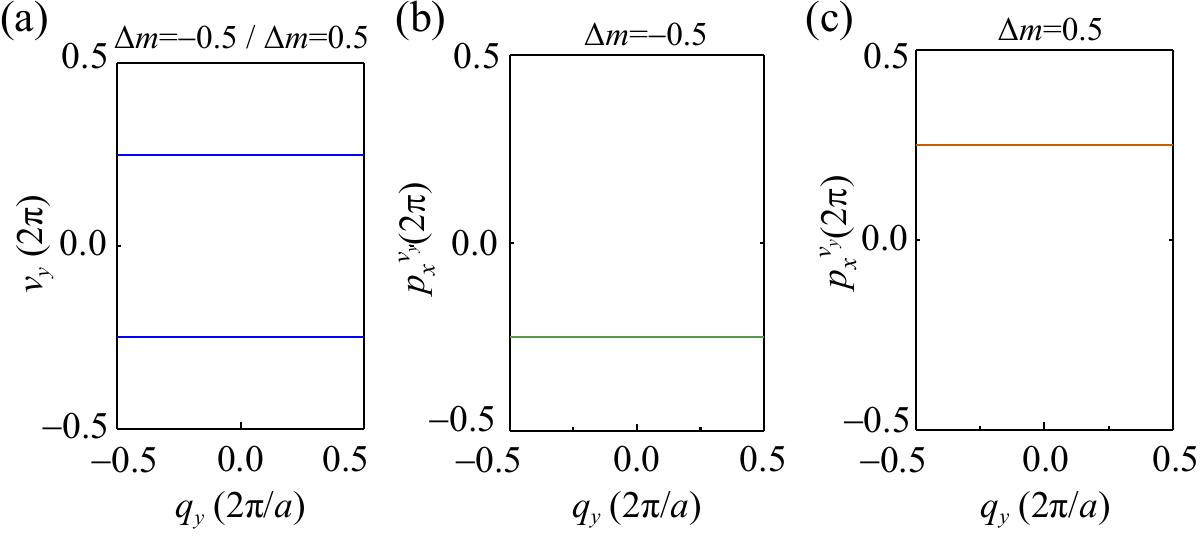}
\caption{\label{fig1b} Calculation of the Wannier band and nested Wannier band: (a) The Wannier band for $\Delta m = -0.5$ and $0.5$, which are identical for both two phases; (b)-(c)The nested Wannier band for $\Delta m = -0.5$ and $0.5$, respectively.   The nested Wannier bands are different and quantized to $-\frac{1}{4}$ and $\frac{1}{4}$. The topological invariant $\Delta p_{x}^{v_{y}}$ is 0.5, which is quantized by the glide symmetry. }
\end{figure}

The 2D square spring-mass lattices are investigated in this section. Various established findings are reinterpreted using topological tools to illustrate and clarify fundamental topology concepts in a straightforward manner. Note that only the out-of-plane motions are taken into account. All coupling flexural stiffness is equal, which is regarded as $k$.  A primitive unit cell shows two masses ($m_a$ and $m_b$) and four springs of constants $k$, as shown in the rhombus box of Fig. \ref{fig1a}(a). Thus,  Fig. \ref{fig1a}(b) and  Fig. \ref{fig1a}(c) denote the corresponding continuum models of the rhombus and rectangle dash box of Fig. \ref{fig1a}(a), respectively.  When mandating $m_a = m_b = 1$, $ k = 1$, and sweeping the wavenumber $\textbf{q}$ along the boundary of the first irreducible Brillouin zone,  a degeneracy line across from high symmetry point $\rm{X_{1}}$ to $\rm{M_{1}}$ is obtained in Fig. \ref{fig1a}(d). To introduce the multiple degenerate states into higher-order topological systems, the extended cell consisting of four masses is considered in the rectangle box of Fig. \ref{fig1a}(a). Then, the quadruple degeneracy solid line determined by the band folding mechanism across from high symmetry point $\rm{M}$ to $\Gamma$ is presented in the solid lines of Fig. \ref{fig1a}(e). In virtue of the glide symmetry $G_{x}: = (x,y) \rightarrow (-x,y+a/2)$ and $G_{y}: = (x,y) \rightarrow (x+a/2,-y)$, inversion symmetry $I: = (x,y) \rightarrow (-x,-y)$, mirror symmetry $M: = (y,x) \rightarrow (-x,-y)$, and translation symmetry $\tau: = (x,y) \rightarrow(x+a/2,y+a/2) $ (see Supplementary Materials), the double degeneracy of the eigenstates on the MX line are protected as  $\Theta_{i}^2 \psi_{n,\textbf{k}} = - \psi_{n,\textbf{k}}$, with $\Theta_{i} = G_{i} \tau (i = x,y)$ \citep{kang2024multiband,zhang2020symmetry}.  $\psi_{n,\textbf{q}}  $ is the Bloch wavefunction, where $n$ represents band index. Unlike the conventional control strategy, we manipulate the topological properties of the unit cell by altering the mass differential. Maintain the mass of the top left and bottom right (top right and bottom left) points equal, the structural dimensionless parameter $\Delta m$ is introduced as follows ${ m_{a} + m_{b} = 2 }, \Delta m = \frac{m_{b}  - m_{a}}{m_{a} + m_{b}}$. The quadruple bandgap is achieved when $\Delta m \neq 0 $, as the solid curves of Fig. \ref{fig1a}(g). Thus, the precise modulation of the mass degree of freedom is of paramount importance, as it dictates the degree of the transition from $C_4$ to $C_2$ symmetry. In this study, we quantify the degree of symmetry breaking through the mass difference, thereby evaluating the evolution of the quadrupole bandgap with $\Delta m \in (-0.5, 0.5)$. As $\Delta m$ transitions from 0 toward  0.5 and $-0.5$, the quadrupole bandgap emerges and reaches its maximum value of 1.2 as shown in Fig. \ref{fig1a}(f). The evolution of the mass and the quadrupole bandgap width exhibits symmetric behavior. The field distributions of these modes for  $\Delta m = 0.5 $ and $= -0.5$ are shown in as shown in Fig. \ref{fig1a}(h), which switch their positions along the frequency axis, revealing the topological phase transition.

To further elucidate topological phase transitions, the (nested) Wannier bands \citep{lin2020anomalous, Benalcazar2017Higher} are employed to characterize the quadrupole topology. The Wilson loop operator along the x-axis is defined for square lattices
\begin{equation}\label{eq12}
		\begin{aligned}
	W_{x,\textbf{k}}(k_{y}) = \mathcal{P} {\rm exp } [\mathcal{I} \oint A_{x,\textbf{q}}dq_{x}]\\ 
\end{aligned}
\end{equation}
where $\mathcal{P}$ denotes the path ording operator one-dimensional Brillouin Zone.  $ \oint A_{x,\textbf{q}}$ denotes the Berry connection (see Supplementary Materials in detail). $x$ represents the direction of the loop. Fig. \ref{fig1b}(a) shows that the Wannier bands are identical for $\Delta m = -0.5$ and $0.5$, and $v_{y}$ are divided into two Wannier sectors. However, the topological properties of the two types of units are still indistinguishable. Since the vanishing dipole moment indicates the quadrupole topology, we can compute the relative quadrupole moments utilizing the nested Wannier band.

 By using the eigenvectors $|v_{x,\textbf{q}}^{j}\rangle$ of $W_{x,\textbf{q}}$, the Wannier band basis can be expressed as $W_{x,\textbf{q}}^{j}\rangle = [v_{x,\textbf{q}}^{j}]^{n} |u_{\textbf{q}}^{n}\rangle $,
where $[v_{x,\textbf{q}}^{j}]^{n}$ denotes the $n$th element of $|v_{x,\textbf{q}}^{j}\rangle$  \citep{Benalcazar2017Higher}. Similar to the definition of the Wilson loop operator, the nested Wilson loop along the y-axis is written as 
\begin{equation}\label{eq18}
\widetilde{W}_{x,\textbf{q}}(q_{y}) = \mathcal{P} {\rm exp } [\mathcal{I} \oint \widetilde{A}_{y,\textbf{q}}dq_{y}]
\end{equation}
where $ \widetilde{A}_{y,\textbf{q}}$ is the Berry connection of Wannier band $v_{x}$.  We diagonalize the nested Wilson loop
\begin{equation}\label{eq21}
\widetilde{W}_{y,\textbf{q}}(q_{x})|p_{y,\textbf{q}}^{v_{x,j}}\rangle = 
	e^{2\pi \mathcal{I}p_{y}^{v_{x,j}}(q_{x})}|p_{y,\textbf{q}}^{v_{x,j}}\rangle
\end{equation}
where $p_{y,\textbf{q}}^{v_{x,j}}$ denotes the polarization for each Wannier band,  defined as $p_{y,j}^{v_{x}} = a\int dq_{x}p_{y,j}^{v_{x}}/2\pi $.  By computing the quadrupole moments, Fig. \ref{fig1b} (b) and (c) show the polarization $p_{y,\textbf{q}}^{v_{x,j}}$ for $\Delta m  = -0.5$ and 0.5 with two distinct Wannier sectors, which are quantized to $-\frac{1}{4}$ and $\frac{1}{4}$, respectively. The relative quadrupole moment is $\frac{1}{4}-(-\frac{1}{4})=\frac{1}{2}$. Thus, the earlier topological phase transition is clarified, also confirming the quadrupole topological features of the bandgap.

\section{Topological transport in continuous square model}

\begin{figure}
\includegraphics[scale=0.375]{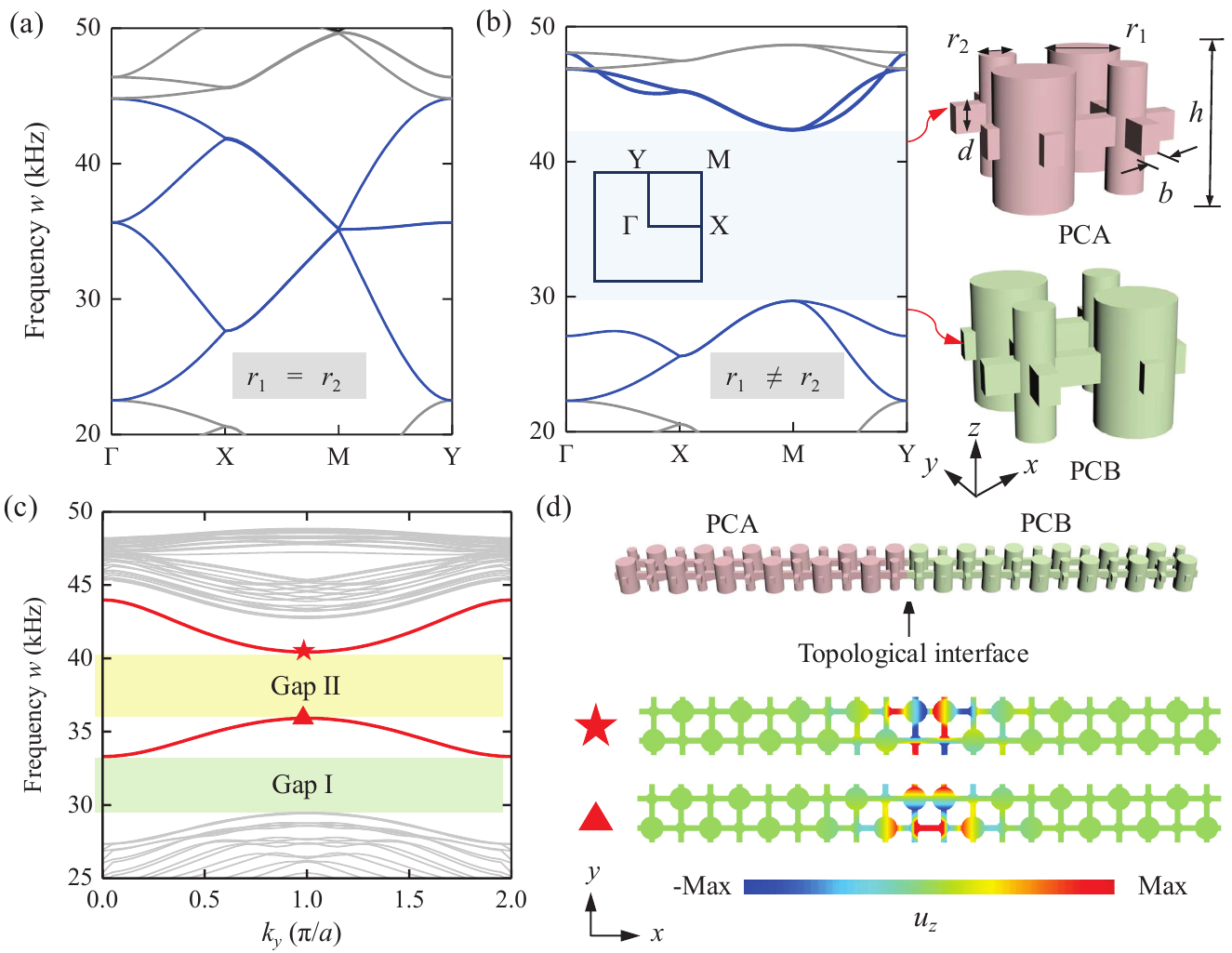}
\caption{\label{fig2}Band structure of the extended unit cell with (a) $r_{1}$ = $r_{2}$ = 2 mm, and (b) $r_{1}$ = 4 mm, $r_{2}$ = 2 mm ($r_{1}$ = 2 mm, $r_{2}$ = 4 mm), in which the illustration on the right is the finite element model of PCA and PCB; (c) Projected band structure of the two-phase strip supercell, the yellow and green areas represent the gap states between the edge states; (d) Finite element model of two-phase strip supercell and modal diagrams of the topological interface corresponding to pentagram and triangle mark, where displacement field is distributed in the center of the interface.}
\end{figure}

\begin{figure}
\includegraphics[scale=0.34]{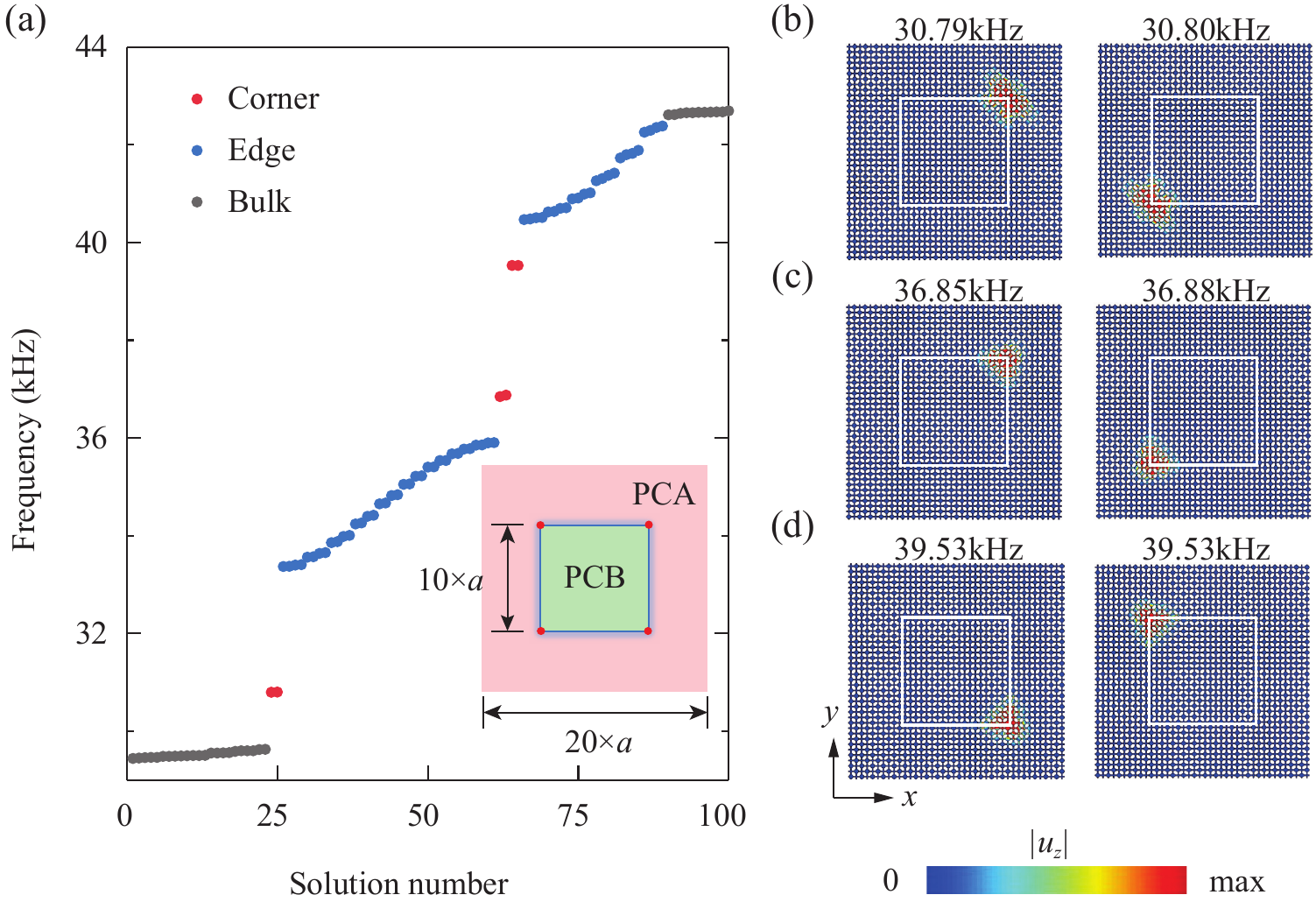}
\caption{\label{fig3}Calculation of the eigenstates for corner configuration: (a) The eigenstates calculation for the square configuration, in which the outer and inner dimensions are 20$a$$\times$20$a$ and 10$a$$\times$10$a$, respectively; three pairs of corner states exist in the gap, which are 30.79 kHz, 36.85 kHz, and 39.53 kHz. The corner, edge, and bulk states are marked in red, blue, and black dots, respectively; (b)-(d) The z-displacement field distribution of the corner states. Two corner states in (d) show a double degeneracy. }
\end{figure}

\begin{figure*}
\includegraphics[scale=0.5]{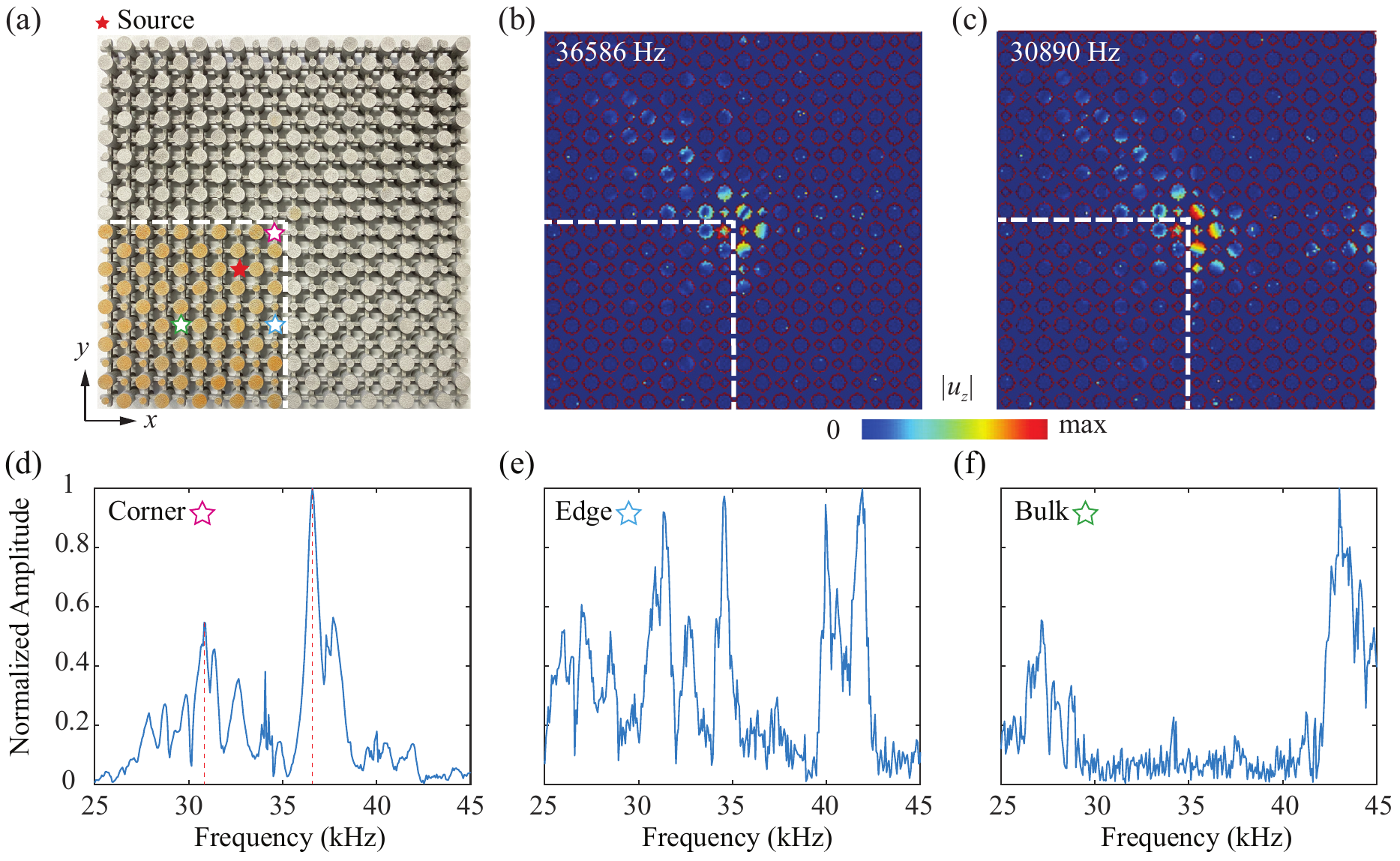}
\caption{\label{fig4}Experimental visualization of the topological corner states: (a) The fabricated sample, where the red pentagram indicates the position of point source excitation. The blue, green, and pink pentagrams represent the detectors for corner, edge, and bulk states,  respectively; (b)-(c) The experimental measured $|u_{z}|$ distribution of two corner states at 36586 Hz and 30890 Hz; (d)-(f) The experimental measured corner, edge, and bulk intensity spectra.}
\end{figure*}

\begin{table} 
	\setlength{\abovecaptionskip}{0.05cm} 
	\centering
	\caption{The initial geometrical parameters of the unit cell}
	\begin{tabular*}{\hsize}{@{\extracolsep{\fill}}c c c c c c} 
		\hline
		$a$/mm & $b$/mm &$h$/mm & $d$/mm & $r_{1}$/mm & $r_{2}$/mm\\
		\hline 
		20 & 2 & 13 & 3 & 3& 3\\		
		\hline
	\end{tabular*}
\label{Table1}
\end{table}

Phononic crystals (PhCs), considered arrays of novel elastic wave material, are widely utilized to explore topological effects as their specific band structure. Hence, we present the PhCs slab made of steel (Young’s modulus $E$ = 206 Gpa, density $\rho$ = 7850 kg/$\rm m^{3}$, and Poisson’s ratio $\mu$ = 0.26) with lattice constant $a$ = 20 mm to simulate the spring-mass model previously mentioned. Fig. \ref{fig2}shows the three-dimensional view of periodic metamaterial,  which denotes all the geometric details of the unit cell. Each unit cell of this system consists of cuboid beams connected with four cylinders, representing the spring stiffness and mass point, respectively. Note that the continuous model in this section realizes the nonsymmorphic symmetries controlled by the radius parameter of the cylinders. The geometric parameters are listed in Table. \ref{Table1}.

By applying periodic boundary conditions and scanning the wave vector $\textbf{q}$, the band structures are obtained in Fig. \ref{fig2}(a) and (b). The unit cell exhibits $C_{4}$ symmetry when the radius parameter of the cylinder fulfills the given relation $r_{1}$ = $r_{2}$, simultaneously resulting in the quadruple degeneracy at high symmetry point M. The band structures highlighted in blue are the ones of interest, distinguished from the gray ones. To produce the bandgap that supports higher-order topology, we alter the radius of the cylinder on the diagonal, which breaks $C_{4}$ symmetry to $C_{2}$ symmetry.  Fig. \ref{fig2}(b) shows the same band structure of the unit cells with  $r_{1}$ = 4 mm, $r_{2}$ = 2 mm (PCA) and $r_{1}$ = 2 mm, $r_{2}$ = 4 mm (PCB), in which the bandgap marked in light blue area ranges from 29.7 kHz to 42.4 kHz. PCA and PCB have been demonstrated to belong to different branches of quadrupole topology. Therefore, the two-phase strip supercell consisting of 12 unit cells is assembled to verify the gapped edge states. Fig. \ref{fig2}(c) also depicts a projected band structure with two edge states. The modal displacement fields at the pentagram and triangle mark are displayed in Fig. \ref{fig2}(d). Besides, the two edge states divide two gaps, which are labeled as Gap $\rm \uppercase\expandafter{\romannumeral1}$ and Gap $\rm \uppercase\expandafter{\romannumeral2}$, respectively.

\begin{figure}
\includegraphics[scale=0.6]{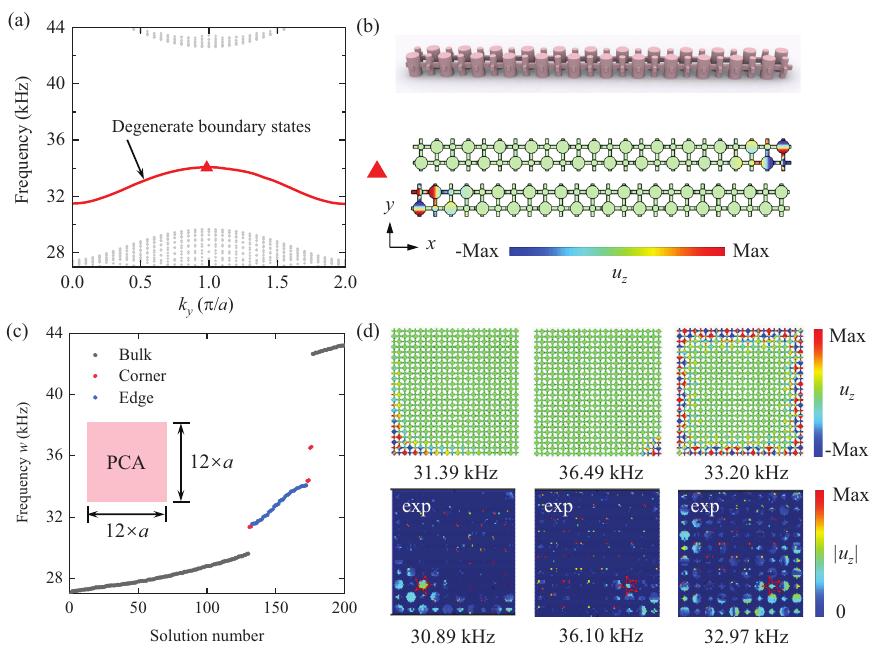}
\caption{\label{fig5}Calculation of the single-phase lattice: (a) The projected band structure of the single-phase strip supercell composed of 12 PCA, where the degenerate boundary states divide the gap between the bulk states into two parts; (b) Finite element model and modal shapes correspond to the boundary states at triangle marks in (a); (c) The eigenstate spectra of the lattice structure composed of 12$\times$12 PCA; (d) The simulated and experimental $|u_{z}|$ distribution of corner states and edge state.}
\end{figure}
\subsection{\label{sec:level2}High topological behavior}

As shown in Fig. \ref{fig3}(a), we construct a square-shaped sample by surrounding the PCB with PCA, in which the outer and inner dimensions are 20$a$$\times$20$a$ and 10$a$$\times$10$a$, respectively. As expected, the numerical eigenstates of this sample confirm that topological corner states are found within the gap beyond the edge states, which are located at 30.79 kHz, 36.85 kHz, and 39.53 kHz. Fig. \ref{fig3}(b)-(d) show the out-of-plane displacement field at corner state frequencies, in which the intensity is highly distributed around corners.

\begin{figure}
\includegraphics[scale=0.33]{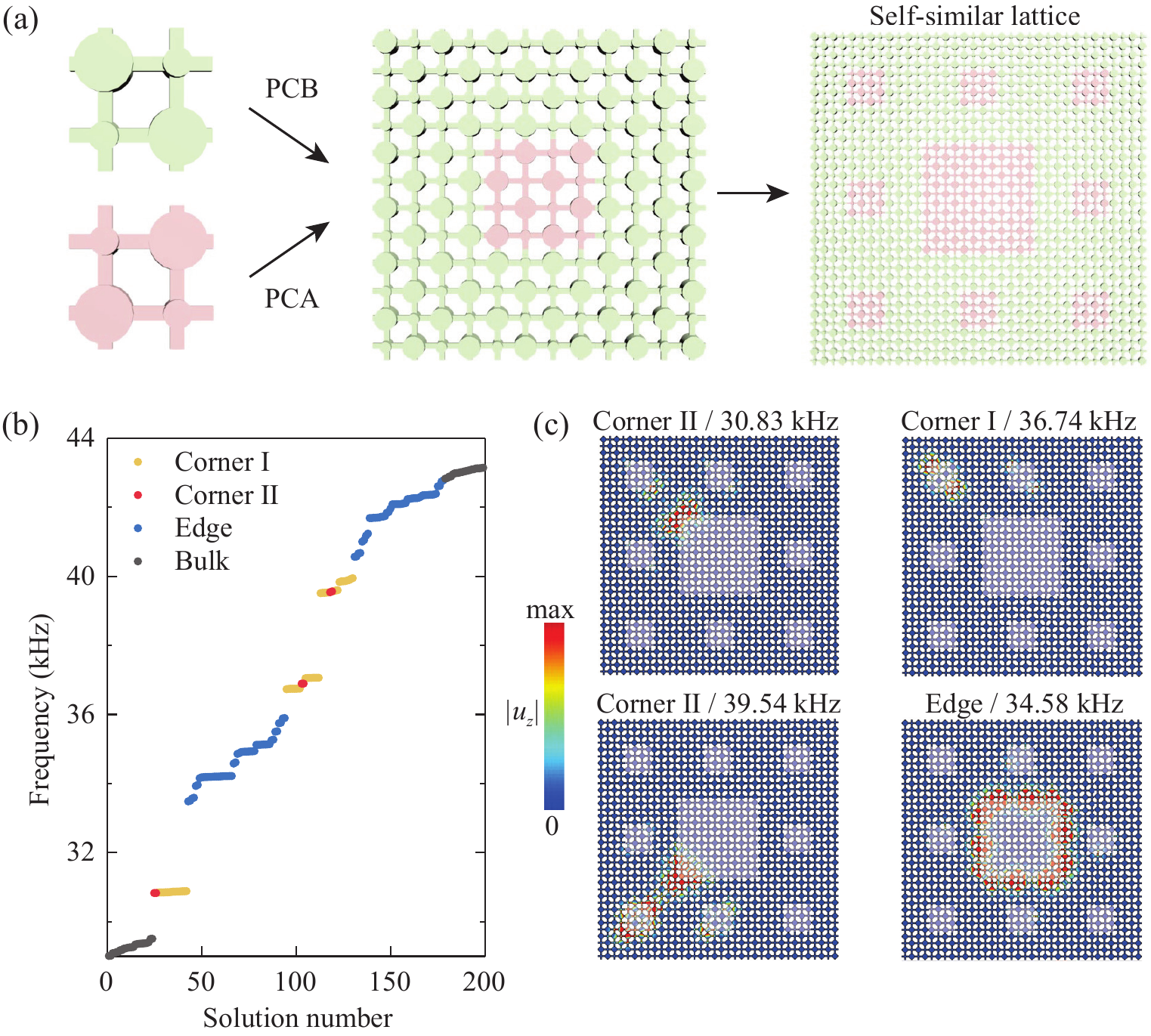}
\caption{\label{fig6}Elastic self-similar lattice: (a) The construction process of the self-similar lattice; (b) The eigenstate spectra of the simulated lattice in (a), where yellow, red, blue, and black dots represent the corner state $\rm \uppercase\expandafter{\romannumeral1}$, corner state $\rm \uppercase\expandafter{\romannumeral2}$, edge, and bulk state, respectively; (c) Schematic of simulated $|u_{z}|$ distribution for the corner state I (36.74 kHz), the corner state $\rm \uppercase\expandafter{\romannumeral2}$ (30.83 kHz, 39.54 kHz), and edge state (34.58 kHz).}
\end{figure}

\begin{figure*}
\includegraphics[scale=0.5]{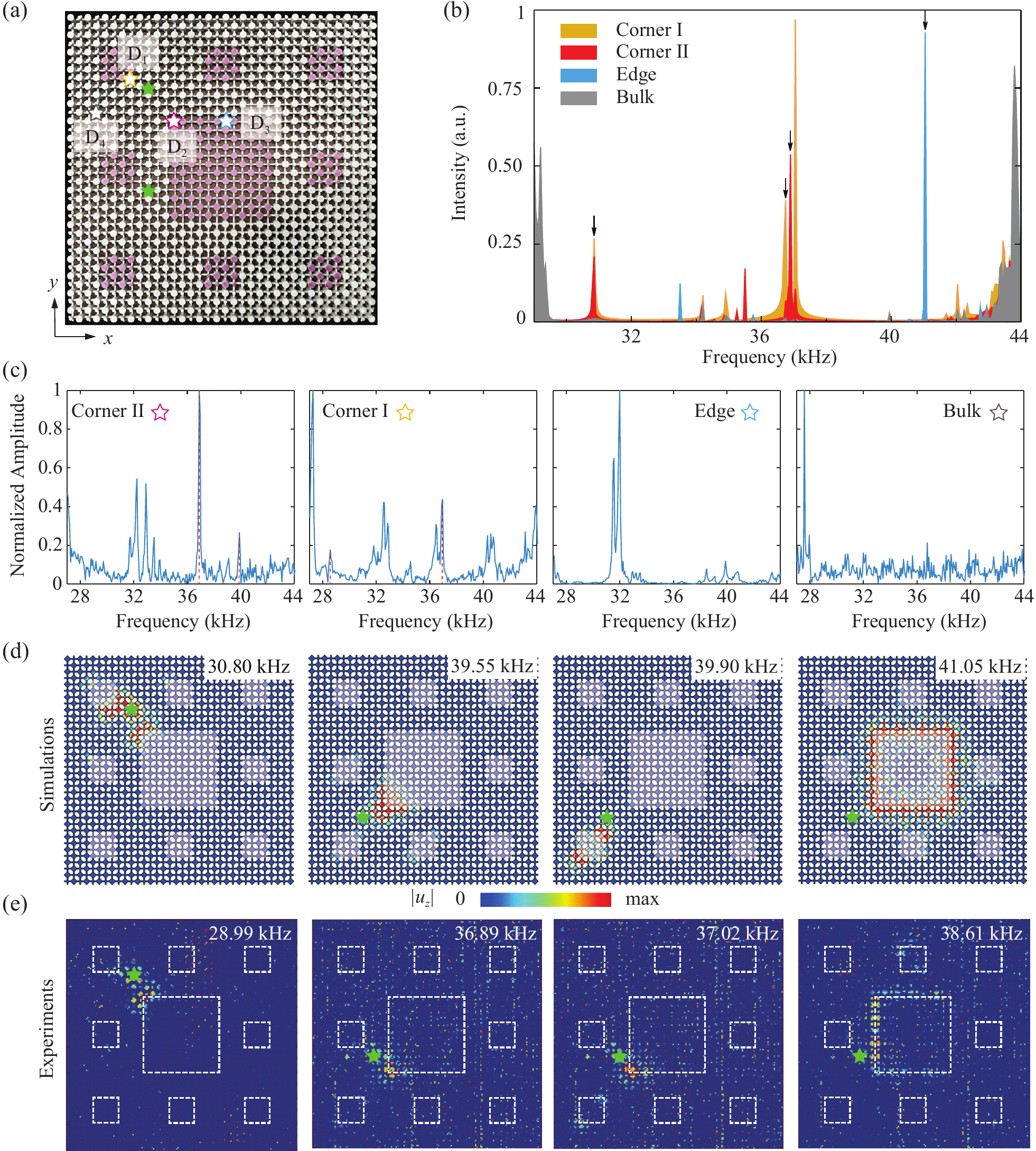}
\caption{\label{fig7}Source-to-detector response spectra of the self-similar lattice: (a) The experimental self-similar sample, where the detectors and point source excitation are set at the on the upper left (or the lower left) of the lattice. The green pentagrams denote the point source; (b) The simulated intensity spectra for four detectors at the position of corner state $\rm \uppercase\expandafter{\romannumeral1}$, corner state $\rm \uppercase\expandafter{\romannumeral2}$, edge, and bulk sites labeled as ``$\rm D_{1}$'', ``$\rm D_{2}$'', ``$\rm D_{3}$'', and ``$\rm D_{4}$'', which are marked by yellow, red, blue, and black, respectively; (c) The experimental measured corner, edge, and bulk intensity spectra; (d) Schematic of simulated $|u_{z}|$ distribution for the corner state II (30.80 kHz), the corner state II (39.55 kHz), the corner state I (39.90 kHz), and edge state (41.05 kHz); (e) The experimental measured $|u_{z}|$ distribution of corner state II (28.99 kHz), corner state II (36.89 kHz), corner state I (37.02 kHz), and edge state (38.61 kHz).}
\end{figure*}

In experiments, the samples are fabricated using the 3D metal-printing technique, like the quarter of the structures in Fig. \ref{fig4}(a).  The model is surrounded by resin to simulate PML. For the topological corner state measurements, we used a piezoelectric ceramic exciter as a point source to generate elastic waves. To prevent interference from reflected waves at the boundaries, we applied ethylene-vinyl acetate copolymers to absorb excess elastic waves. We characterized the amplitude fields by measuring the sample’s frequency responses through spatial scanning, with 1 mm steps along both the x- and y-axes. Amplitudes and phases were captured using a laser vibrometer in combination with a network analyzer. For the transmission curves, we employed a sweep frequency signal as a perturbation to excite the lattice at the input location. The input or output signal was calculated through the surface integral of the total energy flux over a rectangular cross-section containing four sites. The transmission $T(\omega)$ is given by $T(\omega) = W_{output}⁄W_{input}$, where $W$ represents the displacement response of elastic waves. Fig. \ref{fig4}(b)-(c) plots the measured corner, edge, and bulk intensity spectra, respectively. As shown in the measured corner intensity spectra, peaks are located at around 36.58 kHz and 30.89 kHz, respectively. Therefore, we extract the $|u_{z}|$ field distribution at 36.58 kHz and 30.89 kHz, where the experimental results only have $0.72\%$ and $0.32\%$ errors to simulated results (see FIG. S6 in  Supplemental Material). Thus far, it has been demonstrated that all corner states can be activated in the square lattice, presenting the rich topological behavior and possible wave manipulation techniques.

Furthermore, we explore the unique properties of corner states in the single phase of structures supported by quadrupole topology. Through calculating the projected energy band of single-phase strip supercell composed of 12 PCA in Fig. \ref{fig5}(a), the degenerate boundary states represented by red line are obtained, where the boundary states at the triangle mark are given. As shown in Fig. \ref{fig5}(b), only one phase structure performs topological edge states, where the elastic energy mainly concentrates at the end.  Fig. \ref{fig5}(c) shows the eigenstate spectra of structures only in the single phase, whose outer dimension is 12a$\times$12a. In the gaps between the degenerate boundary states and bulk states, the three outer corner states are found at 31.39 kHz, 34.41 kHz, and 36.49 kHz, where Fig. \ref{fig5}(d) shows the experimental $|u_{z}|$ distribution of outer corner states and edge state. It is worth noting that the field distribution on a single site cylinder indicates a dipole mode, as shown in Fig. \ref{fig5}(b). Previous studies \citep{gao2024acoustic} have demonstrated that orbital coupling and the classification of corner states are inextricably interconnected. As shown in the modal shapes of the single-phase supercell, the coupling between the $p_x$ and $p_y$ orbitals can be found in the terminal dimers.  Therefore, the coupling between orthogonal orbitals finally gives rise to anti-symmetric corner states, as shown in Fig. \ref{fig5}(d). In the experiment, the red star indicates the excitation position and the experimental errors to simulated results are $1.59\%$, $1.07\%$, and $0.69\%$, respectively.
 
\subsection{Topological behavior in the self-similarity  mechanical square lattice}

PCA and PCB show topological invariants quantified as -1/4 and 1/4 calculated by the nested Wannier band method. From a spatial point of view, switching between PCA and PCB can be achieved by rotating 90 degrees based on the $xy$ plane. Therefore, in contrast to second-order topological phases \citep{ma2023elastic}, the quadrupole topology can demonstrate the sustained nontrivial nature. To broaden the possibilities of elastic HOTI in more applications, we further investigate the self-similar lattice displayed in Fig. \ref{fig6}(a).  This self-similar lattice is constructed by the minimum primitive cell PCA and PCB, in which the outer and inner dimensions are 6$a$$\times$6$a$ and 2$a$$\times$2$a$, respectively. The self-similar lattice is similar to the Sierpi$\rm \acute{n}$ski carpet. The whole lattice belongs to $\rm G(2)$ and consists of $\rm G(1)$ surrounds. 

The numerical simulation of the self-similar structure is carried out. As shown in Fig. \ref{fig6}(b), in the common gaps, the various eigenmodes are classified as corner state $\rm \uppercase\expandafter{\romannumeral1}$, corner state $\rm \uppercase\expandafter{\romannumeral2}$, edge state, and bulk state. For the corner state $\rm \uppercase\expandafter{\romannumeral1}$ labeled as yellow points, the energy shows a certain aggregation state and gathers at the inner corner points of the smaller blue box, whose $|u_{z}|$ distribution is like the illustration (36.74 kHz) in Fig. \ref{fig6}(c). While the corner state $\rm \uppercase\expandafter{\romannumeral2}$ is labeled as red points, where the energy gathers at the corner points of the central structure completely consisting of PCA, such as the illustration at frequencies of 30.83 kHz and 39.54 kHz. Notably, the eigenstate spectra show similar behavior to the Ref. \cite{LI20222040}. The edge state and the corner state are clearly separated in frequency, while the inner corner states in the center of ${\rm G}(2)$ are surrounded by the inner corner state in ${\rm G}(1)$. Matching the results in Fig. \ref{fig3}, there is no outer corner state in the self-similar lattice. Moreover, 48 corner states $\rm \uppercase\expandafter{\romannumeral1}$ and six corner states $\rm \uppercase\expandafter{\romannumeral2}$ are found, the total numbers correspond to a nine-fold relationship of self-similar structures. It can be observed that both edge states or corner states demonstrate stronger energy localization compared to the fractal lattice. Most importantly, the self-similar lattice, like fractal structures, surpasses the simple periodic structure in terms of number, which potentially excels in sensing, communication, energy harvesting, etc.

Furthermore, the full wave simulation and experimental measurement of the self-similar lattice are carried out with continuous point source excitation input in Fig. \ref{fig7}. As shown in Fig. \ref{fig7}(b), we numerical study the intensity spectra for four detectors at the position of corner state $\rm \uppercase\expandafter{\romannumeral1}$, corner state $\rm \uppercase\expandafter{\romannumeral2}$, edge, and bulk sites labeled as ``$\rm D_{1}$'', ``$\rm D_{2}$'', ``$\rm D_{3}$'', and ``$\rm D_{4}$'' in Fig. \ref{fig7}(a). The green pentagram indicates the point source excitation. The bulk curve produces a response at both ends of the spectrum, which is consistent with the description in Fig. \ref{fig6}(b). The corner state $\rm \uppercase\expandafter{\romannumeral1}$ (yellow curve) and corner state $\rm \uppercase\expandafter{\romannumeral2}$ (red curve) are distributed in the common gap, where the red curve has one peak located at 39.55 kHz. Note that the number and frequency of corner state $\rm \uppercase\expandafter{\romannumeral2}$ in the self-similar structure are consistent with corner states at the upper right of the periodic structure, which is attributed to the opposite construction of corner state configuration. We can clearly find in Fig. \ref{fig7}(b), that the peaks between corner states $\rm \uppercase\expandafter{\romannumeral1}$ and $\rm \uppercase\expandafter{\romannumeral2}$ are separated from each other, and there is only one peak seen in the curve of corner states $\rm \uppercase\expandafter{\romannumeral1}$ and corner states $\rm \uppercase\expandafter{\romannumeral2}$, respectively. When moving the excitation source to the lower left of the sample in the experiment, Fig. \ref{fig7}(c) plots the measured intensity spectra. Fig. \ref{fig7}(d) shows the simulated $|u_{z}|$ distribution for the corner state I (30.80 kHz), the corner state II (39.55 kHz), the corner state I (39.90 kHz), and edge state (41.05 kHz), respectively. Fig. \ref{fig7}(e) shows the measured $|u_{z}|$ field distribution diagram of corner state I (28.99 kHz), corner state II (36.89 kHz), corner state I (37.02 kHz), and edge state (38.61 kHz), respectively. Compared with simulation results in Fig. \ref{fig7}(d), the experimental errors of corner state $\rm \uppercase\expandafter{\romannumeral1}$ and corner state $\rm \uppercase\expandafter{\romannumeral2}$ are $5.87\%$, $7.21\%$ and $6.73\%$, respectively. We can observe the error in the self-similar structure is larger than that in the simple periodic structure, which can be attributed to accumulated manufacturing error with size expansion.
\section{Conclusion} 
HOTIs provide a promising approach to realizing and controlling corner states. Previous research on higher-order topology has frequently relied on negative coupling, which is relatively easy to implement in quantum systems but presents significant challenges in classical systems. For instance, Qi et al. \citep{qi2020acoustic} developed a specialized design for airborne sound to achieve negative coupling and realize higher-order topology. However, elastic HOTI with quadrupole topology, without the need for negative coupling, has yet to be thoroughly explored. Our work shows that quadrupole topology can be realized in elastic systems without negative coupling. 

In this work, we have thoroughly explored the topological behavior of quadrupole HOTI in elastic lattice through theoretical, numerical, and experimental approaches. By applying the spring-mass theory to a square lattice, we predict the presence of three pairs of corner states within the phononic band gaps (see Supplement Material). Using topological tools such as the nested Wannier band, we identify the system as exhibiting nontrivial quadrupole topology, and the corner states are derived from the superlattice structure. These topological corner states can be successfully excited by numerical and experimental methods. Meanwhile, the frequencies of the topological states in the numerical models are almost identical to the experimental results. Additionally, we explore the edge and corner states in a self-similar lattice, demonstrating that corner states can exist at different lattice sites and even couple at the same frequency. Given that the relative quadrupole behaviors in the elastic lattice are effectively predicted by the spring-mass models and validated through consistent simulations and experiments, this work offers fresh insights that may inspire the development of a diverse array of new topological materials.

\begin{acknowledgments}
The authors wish to acknowledge the support from the National Nature Science Foundation of China (Grant Nos. 12002094, 12272040, 12374157, and 12302112), and the Fundamental Research Funds for the Central Universities (No. 30923010207).
\end{acknowledgments}

\nocite{*}

\bibliography{apssamp}

\end{document}